\documentclass[aps,prl,twocolumn,superscriptaddress,longbibliography]{revtex4-2}

\usepackage{setspace}
\usepackage{amsmath}
\usepackage{lipsum}  
\usepackage{graphicx}
\usepackage{verbatim}
\usepackage{subfigure}
\usepackage[english]{babel}
\usepackage{verbatim}
\usepackage{multirow}
\usepackage{xcolor}
\usepackage[utf8]{inputenc}
\usepackage{booktabs}
\usepackage{dcolumn}
\usepackage{bm}
\usepackage[flushleft]{threeparttable}
\usepackage{graphics}
\usepackage{afterpage}
\usepackage{float}

\newcolumntype{.}{D{.}{.}{-1}}
\newcolumntype{d}[1]{D{.}{.}{#1}}

\usepackage[colorlinks,linkcolor=blue,anchorcolor=blue,citecolor=blue,]{hyperref}

\usepackage{lipsum}
\usepackage{siunitx}

\begin{document}

\title{The {Z}ero-{Q}uantum-{D}efect Method and the {F}undamental {V}ibrational {I}nterval of H$_{2}^+$}
\author{I. Doran}
\affiliation{Departement of Chemistry and Applied Biosciences, ETH Zurich, Zurich, Switzerland}
\author{N. H\"{o}lsch}
\affiliation{Departement of Chemistry and Applied Biosciences, ETH Zurich, Zurich, Switzerland}
\author{M. Beyer}
\affiliation{%
	Department of Physics and Astronomy, LaserLaB, Vrije Universiteit Amsterdam, de Boelelaan 1081, 1081 HV Amsterdam, The Netherlands%
}
\author{F. Merkt} 
\email{frederic.merkt@phys.chem.ethz.ch (F. Merkt)} 
\email[]{merkt@phys.chem.ethz.ch}
\affiliation{Departement of Chemistry and Applied Biosciences, ETH Zurich, Zurich, Switzerland}
\affiliation{Departement of Physics, ETH Zurich, Zurich, Switzerland}
\affiliation{Quantum Center, ETH Zurich, Zurich, Switzerland}

\date{\today}

\begin{abstract}
The fundamental vibrational interval of H$_{2}^+$ has been determined to be $\Delta G _{1/2} = 2191.126\,614(17)$ cm$^{-1}$ by continuous-wave laser spectroscopy of Stark manifolds of Rydberg states of H$_2$ with the H$_{2}^+$ ion core in the ground and first vibrationally excited states.  Extrapolation of the Stark shifts to zero field yields the zero-quantum-defect positions $-R_{\textrm{H}_2}$/$n^2$, from which ionization energies can be determined. Our new result represents a four-order-of-magnitude improvement compared to earlier measurements. It agrees, within the experimental uncertainty, with the value of 2191.126\,626\,344(17)(100) cm$^{-1}$ determined in non-relativistic quantum electrodynamic calculations [\href{http://doi.org/10.1103/PhysRevLett.118.233001}{V. Korobov, L. Hilico and J.-Ph. Karr, Phys. Rev. Lett. 118, 233001 (2017)}].
\end{abstract}

\maketitle

H$_{2}^+$ is a fundamental molecular three-body quantum system. Its energy-level structure can be calculated with high accuracy in first-principles quantum-mechanical calculations including relativistic and quantum-electrodynamics (QED) corrections \cite{bishop77a, moss93a, korobov06a, korobov08a, korobov17a}. In a calculation of QED corrections up to order $m \alpha^7$, even including the largest correction of order $m \alpha^8$, Korobov et al. determined the fundamental vibrational frequency $\Delta G _{1/2}$ in H$_{2}^+$ to be 65\,688\,323\,710.1(5)(2.9) kHz (2191.126\,626\,344(17)(100) cm$^{-1}$) \cite{korobov17a}, where the first and second uncertainties are the theoretical uncertainty and the uncertainty caused by the value of the proton-to-electron mass ratio recommended at that time \cite{mohr16a}. H$_{2}^+$ does not have a permanent electric-dipole moment and precision measurements of this interval are challenging. The most precise experimental values for the fundamental vibrational interval to date remain the value of 2191.2(2) cm $^{-1}$ determined by Herzberg and Jungen in 1972 \cite{herzberg72a} through extrapolation of the Rydberg series observed in the absorption spectrum of H$_2$ and the value of 2191.126\,53(8) cm$^{-1}$ reported in the dissertation of Beyer \cite{beyer18e}, also obtained by Rydberg-series extrapolation. Efforts are currently underway in several laboratories to measure rovibrational intervals in H$_{2}^+$ using cold H$_{2}^+$ ions in ion traps \cite{karr16a, schmidt20a, schwegler23a, alighanbari23a}.

HD$^+$ has an electric-dipole moment resulting from the displacement of the center of mass from the center of charge, which facilitates precision measurements of rovibrational transitions. Recent experiments using sympathetically cooled HD$^+$ ions have resulted in accurate transition frequencies \cite{patra20a, alighanbari20a, kortunov21a}, which are used in the determination of the proton-to-electron mass ratio and other physical constants \cite{karr16a, alighanbari20a, germann21a}. However, discrepancies concerning the hyperfine structure remain unexplained \cite{koelemeij22a, karr23a} and measurements in H$_{2}^+$ have thus regained attractivity. 
\begin{figure}[h]\centering
	{\includegraphics[trim=0cm 0.0cm 0cm 0cm, clip=false, width=0.8\linewidth]{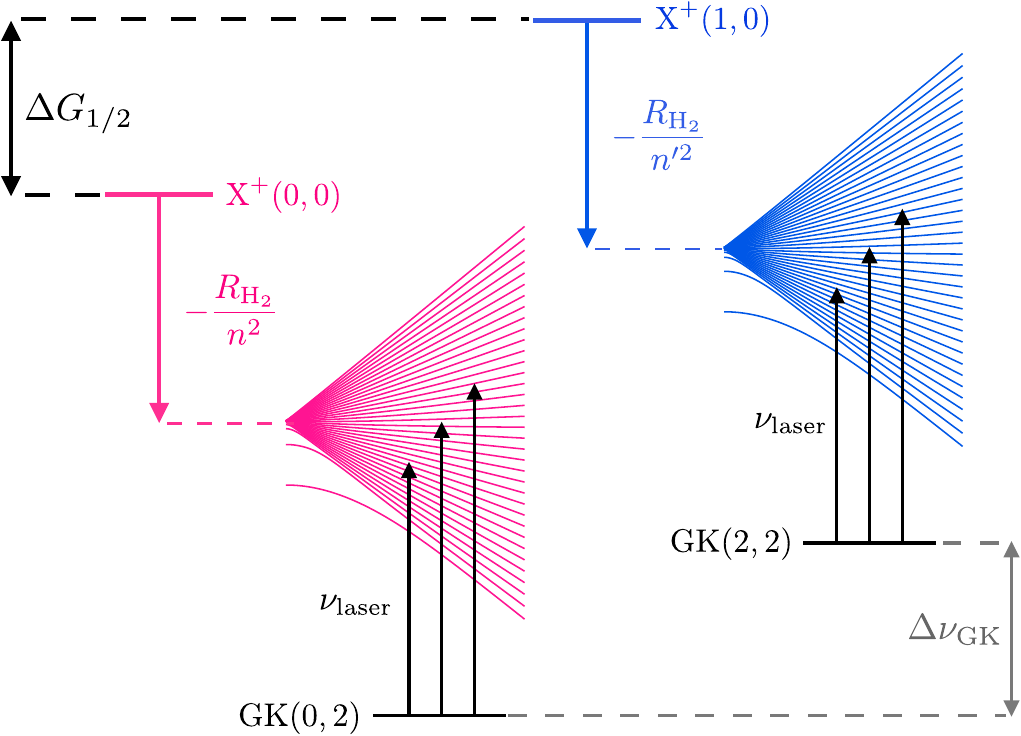}}
      \caption{Energy-level diagram and intervals used to determine the fundamental vibrational interval of H$_2^+$ ($\Delta G _{1/2}$). Transitions from the two GK($v$,2)  intermediate states to the Rydberg-Stark manifolds with $v^+=0$ (magenta) and $v^+=1$ (blue) ion cores are shown as black arrows ($\nu_{\textrm{laser}}$). The zero-quantum-defect positions ($-R_{\textrm{H}_2}/n^{2}$) are shown with respect to the $v^+=0$ and $v^+=1$ ionization thresholds. The relative energy spacings are not to scale.}  
              \label{fig:diagram}	
\end{figure}
We present a determination of the fundamental vibrational interval of H$_{2}^+$ from measurements in high-$n$ Rydberg states of H$_2$ following the scheme depicted in Fig. \ref{fig:diagram}. The measurements rely on the derivation of the positions of zero-quantum defect (ZQD), i.e., of the Bohr energies $-R_{\textrm{H}_2}$/$n^2$, where $n$ represents the principal quantum number and $R_{\textrm{H}_2}$ the mass-corrected Rydberg constant for H$_2$. These positions are obtained from the analysis of the Stark shifts of high Rydberg states of H$_{2}$ in weak electric fields. This method, referred to as the ZQD method, was introduced in a millimeter-wave spectroscopic measurement \cite{hoelsch22a} of the Stark effect in high-$n$ Rydberg states of H$_2$ with an X$^+ \, ^{2} \Sigma_g^+$ ($v^+=0, N^+=0$) ion core. We extend it here to the Rydberg series converging to the $v^+=1, N^+=0$ state of H$_{2}^+$ which we measure from selected rovibrational levels of the GK $^1\Sigma_g^+$ state of H$_{2}$.

The ZQD method relies on the precise calculation of the binding energies of Rydberg-Stark states. In summary, the Hamiltonian matrix corresponding to $\hat{H}_0 + eF\hat{z}$ is constructed in the Hund's case d) basis set $| n \ell N^+NM_N \rangle $. The matrix $\hat{H}_0 $ is diagonal and its elements are determined either from multichannel-quantum-defect-theory (MQDT) calculations for states with $\ell \leq 5$ \cite{jungen11a, osterwalder04a, sprecher14a} or from a polarization model for states with $\ell > 5$ \cite{eyler83a}.  We also use experimentally determined zero-field energies for the p and d states, which, in para-H$_2$, are not predicted by MQDT as accurately as $\ell \geq 3$ states (see Supplemental Material for an example).

\begin{figure}[h]
	\includegraphics[trim=0cm 0cm 0cm 0cm, clip=false, width=\linewidth]{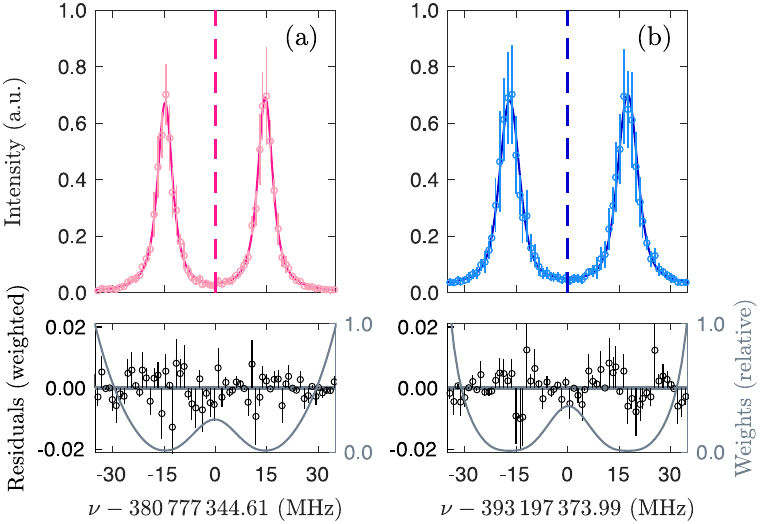}
	\caption{Upper panels: spectra of the 70f0$_3$ ($v^+$=0) $\leftarrow$ GK(0,2) (a) and 70f0$_3$ ($v^+$=1) $\leftarrow$ GK(2,2) (b) transitions (dots) and fits using a Voigt line-shape model (solid lines). The vertical dashed lines indicate the Doppler-free transition frequencies. Lower panels: weighted residuals and their corresponding relative weights (gray traces).}
	\label{fig:spectraff}
\end{figure}

The experiments are carried out using the same procedures and apparatus as described in Refs. \cite{beyer18a, hoelsch22a}. They involve a pulsed skimmed supersonic beam of pure H$_2$ (beam velocity $v$ $\approx$ 1200 m/s) emitted from a valve held at 60 K into a high-vacuum chamber. The H$_2$ molecules are excited to high-$n$ Rydberg states through the resonant three-photon excitation sequence:
\begin{align}
      \text{X}\,^1\Sigma_g^+(0,0)  \xrightarrow{\mathrm{VUV}} &~\text{B}\,^1\Sigma_u^+(4,1) \notag\\
       \xrightarrow{\mathrm{VIS}} &~\text{GK}\,^1\Sigma_g^+(0,2)~\text{or}~(2,2)  \label{eq:excischeme} \\
       \xrightarrow{\mathrm{NIR}} &~n\mathrm{p} /n\mathrm{f} \,\,[\text{H}_2^+\,\text{X}{^{+}}\,^{2} \Sigma_g^+ (0,0) \  \text{or} \ (1,0)]\,,\notag
\end{align}
where the numbers in parentheses, ($v,N$) for H$_2$ and ($v^+, N^+$) for H$_{2}^+$, indicate the vibrational quantum number and the quantum number for the total angular momentum without spin. The frequency interval $\Delta \nu_{\textrm{GK}}$ between the GK(0,2) and GK(2,2) intermediate states is accurately known \cite{hoelsch18a} and used to determine $\Delta G_{1/2}$ according to (see also Fig. \ref{fig:diagram}):
\begin{align}
    hc \Delta G_{1/2}  &=  E_\textrm{I}[ v^+ = 1 \leftarrow \textrm{GK}(2,2)] \ - \notag\\
    & \ \ \ \ E_\textrm{I} [ v^+ = 0 \leftarrow \textrm{GK}(0,2)] \ + hc \Delta \nu_{\textrm{GK}}\,.
\label{eq:Enint}
\end{align}
The pulsed (repetition rate 25 Hz) vacuum-ultraviolet (VUV) and visible (VIS) lasers used to populate the GK(0,2) and (2,2) intermediate states are described in Ref. \cite{beyer18a}. Transitions to Rydberg states are induced from these levels using a single-mode cw near-infrared (NIR) laser (bandwidth 1 MHz), which crosses the supersonic beam at near-right angles, with a small deviation angle $\beta$ from 90$^\circ$. To determine and eliminate Doppler shifts, the NIR laser is retroreflected by a mirror located beyond the photoexcitation region. This configuration leads to two Doppler components for each transition, with opposite first-order Doppler shifts $ \pm \frac{\textrm{$v$\,sin}\beta } {\textrm{$c$}} \cdot \nu_{\textrm{NIR}} $ (see Fig. \ref{fig:spectraff}). The center frequency is determined by averaging their frequencies, after overlapping the incoming and reflected laser beams at a distance of 12 m (see Ref. \cite{beyer18a} for details). This procedure automatically eliminates any contribution to the Doppler shifts that could arise from the selection of certain velocity classes through the resonant two-photon excitation from the X(0,0) ground state to the GK $^1\Sigma_g^+(v,2)$ intermediate states.

Fig. \ref{fig:spectraff}a) and b) displays spectra measured under field-free conditions for the two transitions 70f0$_3$ ($v^+$=0) $\leftarrow$ GK(0,2) and 70f0$_3$ ($v^+$=1) $\leftarrow$ GK(2,2). Individual Doppler components have linewidths of 5.2(5) and 6.9(7) MHz, respectively. The slight broadening of the 70f0$_3$ ($v^+$=1) $\leftarrow$ GK(2,2) transition is attributed to vibrational autoionization. The measurements of Stark spectra were performed with one Doppler component only, to avoid spectral congestion. To correct for the first-order Doppler shift, two field-free Doppler-compensation spectra were recorded, one at the beginning and one at the end of the measurement session, and it was verified that they yielded identical values of the Doppler shifts.

The NIR laser frequency is calibrated to an accuracy of 2 $\times$ 10$^{-11}$ ($\Delta \nu/ \nu$) using a frequency comb which is referenced to a Rb oscillator disciplined by a GPS receiver \cite{beyer18a}.

Magnetic stray fields in the photoexcitation volume are reduced to below 1 mG using a double-layer mumetal magnetic shield. To both compensate stray electric fields for the measurement of zero-field spectra and impose well-controlled electric fields (typically below 1.5 V/cm) for the measurement of Stark spectra, dc potentials are applied across an electrode stack surrounding the photoexcitation region. Field-ionization of long-lived Rydberg states below the $v^+=0, N^+=0$ ionization threshold is achieved by applying large pulsed potentials across the stack. The resulting electric fields extract the H$_2^+$ ions, including those generated by autoionization above the $v^+=0$ ionization threshold, towards a microchannel-plate detector. Spectra are recorded by monitoring the H$_2^+$ ion signal as a function of the NIR laser frequency. 

Measurements of Rydberg-Stark manifolds in series converging to the $v^+=0$ and 1 H$_2^+$ thresholds were performed at several values of $n$ between 45 and 70, as summarized in Fig. \ref{fig:resistark} below.
For each selected $n$ value, spectra were recorded at several electric field strengths in the range between 90 and 300 mV/cm, where the Stark effect of the nearly degenerate high-$\ell$ ($\ell \geq 3$) states, called high-$\ell$ manifold below, is essentially linear. The polarizations of all lasers were parallel to the applied electric field, which, when $N^+=0$, restricts the excitation to $m_{\ell} = 0$ Rydberg-Stark states. The intermediate GK(0,2) and (2,2) states have dominant 3d$\sigma$ character. Consequently, transitions are to Rydberg states with $n$p or $n$f character. Because transitions to $n$p Rydberg states were found to be about two orders of magnitude weaker than to $n$f states, line intensities directly reflect the $\ell=3$ character of each $k$ state. In the calculation of intensities, we therefore assume them to be proportional to the f character of the Stark states.

\begin{figure*}[ht]
	\begingroup
	\fboxsep=0mm
	\fboxrule=0.0pt
		\fbox{\includegraphics[trim=0cm 0cm 1.0cm 0cm, clip=false, width=0.98\linewidth]{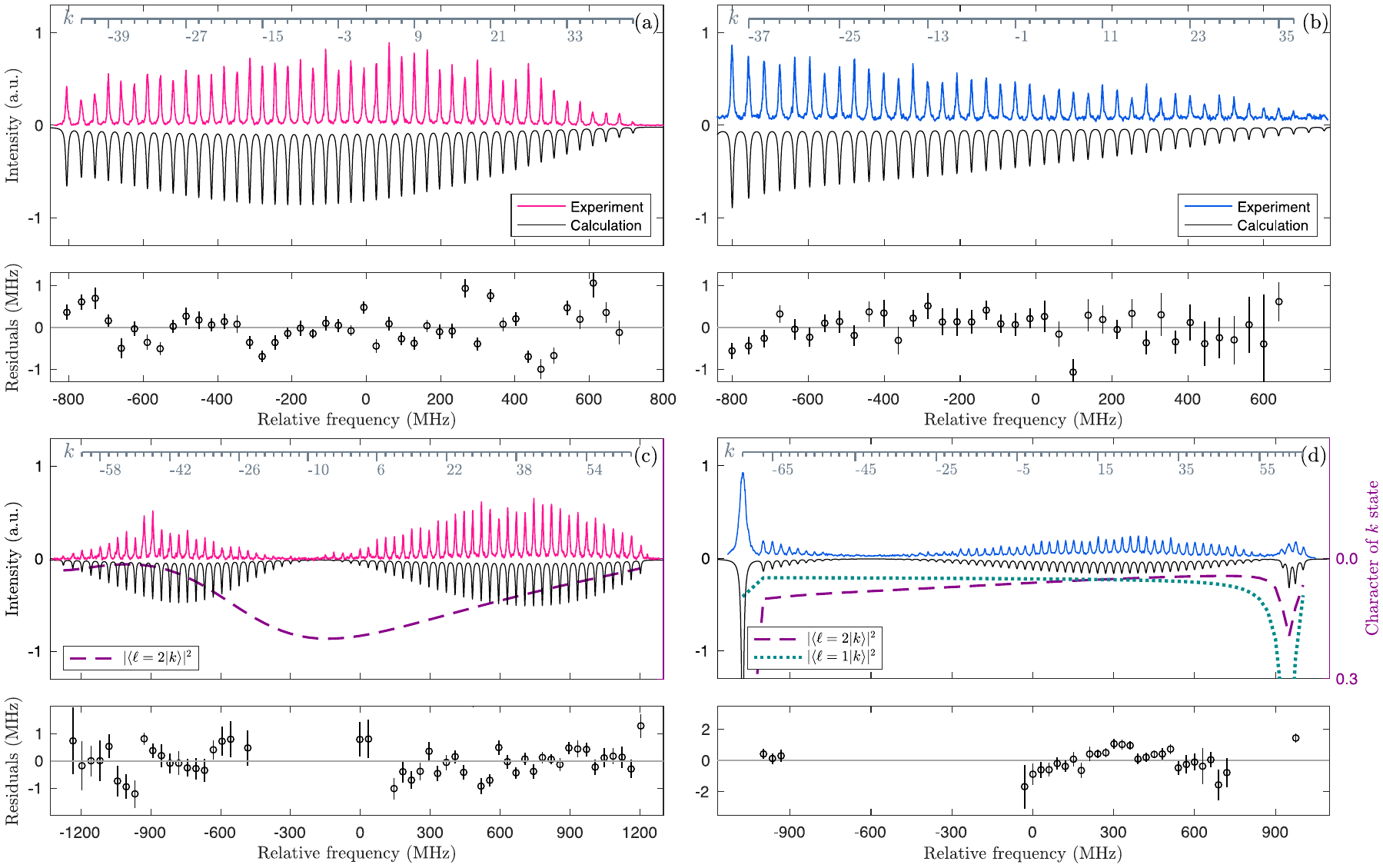}}
	\endgroup
	\caption{Spectra of the $n$ = 48 [X$^+$(0,0)] (a, magenta trace), $n$ = 46 [X$^+$(1,0)]  (b, blue trace), $n$ = 69 [X$^+$(0,0)]  (c, magenta trace) and $n$ = 70 [X$^+$(1,0)]  (d, blue trace) Rydberg-Stark manifolds and calculations (black traces), recorded at values of the electric fields of 180 mV/cm (a), 210 mV/cm (b), 140 mV/cm (c) and 110 mV/cm (d). The purple (turquoise) dashed (dotted) lines in (c) and (d) indicate the relative $| \langle \ell = 2 | k \rangle | ^2$ ($| \langle \ell = 1 | k \rangle | ^2$) character of each $k$ state. The $k$ quantum numbers assigned to each Rydberg-Stark state are indicated on the grey axes. The lower panels show the residuals of the fit of the Stark calculation to the experimental line positions. The error bars indicate the uncertainties in the determination of the experimental line positions. All frequencies are referenced to the ZQD position for each value of $n$.}
	\label{fig:spectrastark}
\end{figure*}
Fig. \ref{fig:spectrastark} shows spectra of X$^+$(0,0) [$n$ = 48 (a), 69 (c)] and X$^+$(1,0) [$n$ = 46 (b), 70 (d)] Rydberg-Stark manifolds. The Rydberg-Stark states are labeled by the integer number $k$, which varies between $-(n-1-|m_{\ell}|) $ and $(n-1-|m_{\ell}|)$ in steps of two, in order of increasing energy \cite{bethe57a}. Fig. \ref{fig:spectrastark}a) and b) exhibits regular and smoothly varying intensity patterns resulting from the homogeneous distribution of the f character among the different $k$ states. At these $n$ values, the $\ell=0$--2 states have large quantum defects and are located outside of, and are not significantly mixed with, the high-$\ell$ Stark manifold. The Stark states therefore exclusively have $\ell \geq 3$ character and autoionization is inhibited by the field-induced mixing of nonpenetrating high-$\ell$ character. The calculated spectra (inverted black traces) are in excellent agreement with the measured ones.

When the $\ell = 0$--2 states have a small enough quantum defect, they are already fully mixed in the high-$\ell$ Stark manifolds at low fields. In this case, the intensity distributions become irregular and exhibit minima located near the zero-field positions of the low-$\ell$ states, where the f character is reduced, as illustrated in Fig. \ref{fig:spectrastark}c) and d). In the case of $v^+=0, n = 69$ [Fig. \ref{fig:spectrastark}c)], the s and p Rydberg states are located far from the manifold, but the quantum defect of the d state is small enough for  this state to be fully integrated in the Stark manifold. The intensity minimum is located where the d character (dashed purple line) is maximal and the $\ell=3$ character minimal. This behavior is accounted for by the calculations of relative intensities (black inverted spectrum). For the $v^+=1, n = 70$ state [Fig. \ref{fig:spectrastark}d)], both the p and d Rydberg states have small enough quantum defects to be integrated in the Stark manifold. Two minima result in the intensity distribution, which deviate from the calculated positions of maximal p (dotted turquoise line) and d (dashed purple line) character because of the field-induced interaction between these states.

\begin{figure}
	{\includegraphics[trim=0cm 0cm 0cm 0cm, clip=false, width=0.98\linewidth]{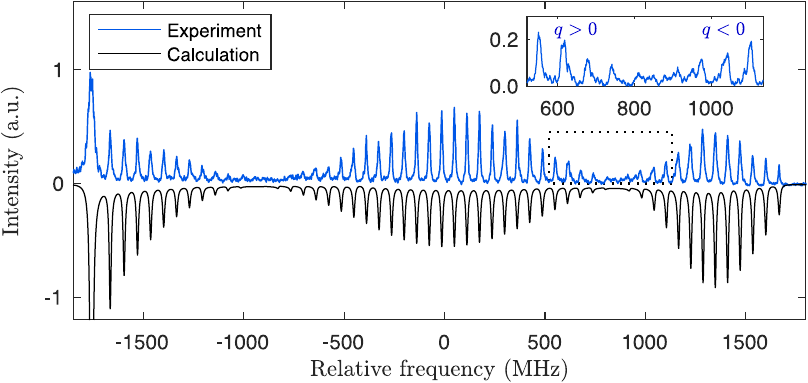}}
      \caption{Spectrum of the $n$ = 56 [X$^+$(1,0)] Rydberg-Stark manifold recorded from the GK(2,2) state at a value of the electric fields of 300 mV/cm (blue trace), and calculation (black trace). The inset shows an expanded version of the spectrum corresponding to the region indicated by the dotted black rectangle. All frequencies are referenced to the ZQD position for $n$ = 56.}  
              \label{fig:fanoshape}
\end{figure}

The admixture of core-penetrating low-$\ell$ character enhances the autoionization of the $v^+=1$ Rydberg-Stark states, which leads to spectral broadening and to asymmetric lineshapes (Fano profiles \cite{fano61a}) in the regions near the intensity minima. An example is provided in Fig. \ref{fig:fanoshape}, which shows the $v^+=1, n = 56$ Stark manifold. This spectrum reveals two regions of weak intensities originating from the small quantum defects of the 56p0$_1$ and 56d0$_2$ Rydberg states. In these regions, the lines display asymmetric Fano profiles, with a change in sign of the Fano parameter $q$ when passing through the positions of minimal intensity. This \textit{q}-reversal phenomenon has been previously observed in studies of atomic and molecular Rydberg states (see Ref.~\cite{jungen81a, gounand83a, du86a, kung86a, domke91a, kim93a, greetham03a, viteri07b, wang17a}) and is attributed to complex resonances involving the interaction of at least two closed and one open channel \cite{jungen81a, gounand83a, du86a, kung86a, domke91a, kim93a, greetham03a, viteri07b}. Here, we observe a $q$ reversal when $\ell <3$ states are mixed into the $v^+=1$ high-$\ell$ Rydberg-Stark manifolds. The complex resonances result from the interaction between (i) the Rydberg-Stark states with predominantly high-$\ell$ character that are only very weakly coupled to the $v^+=0$ continua, (ii) the low-$\ell$ states that are strongly coupled to both the high-$\ell$ Rydberg-Stark states and the $v^+=0$ continua, and (iii) the $v^+=0$ continua. In the perturbed regions, the experimental lineshapes were analyzed using Fano profiles of adjustable widths, line centers and $q$ parameter to determine reliable positions (see inverted calculated spectrum in Fig.~\ref{fig:fanoshape}).

The calculated Stark-state energies were matched to the observed relative positions by optimizing the field strength and locating the ZQD position in the measured spectra, as illustrated schematically in Fig. \ref{fig:diagram}, in a least-squares fit. The ionization energies were obtained by adding $R_{\textrm{H}_2}/n^{2}$. The residuals between calculated and measured line positions are typically below 1 MHz (see lower panels of Fig. \ref{fig:spectrastark}).

\begin{table}[h]
\caption{Error budget and frequency corrections for the determination of the binding energy of the GK(0,2) [GK(2,2)] state from the measurement of transitions to the Rydberg-Stark manifold for one value of $n$, and one value of the electric field strength. All values and uncertainties are reported in kHz.}
\label{tab:errtr}
\resizebox{3.3in}{!} {
\begin{tabular}{llll} 
\toprule
 & $\Delta \nu$  &  $\sigma _\textrm{stat}$  & $\sigma _\textrm{sys}$ \\
\midrule
Least-squares fit of ZQD method & &  $<$500$^\textrm{a}$& \\
Residual 1$^ {\textrm{st}}$ order Doppler shift & &  & 250$^\textrm{b}$\\
Line-shape model & & \ \ \ \ & 100(200)$^\textrm{c}$ \\
2$^ {\textrm{nd}}$ order Doppler shift & \ \ +2 & & 0.5 \\
ac Stark shift &  & &$\sim$5 \\
Zeeman shift &  &  &$\sim$10 \\
Pressure shift &  &  &$\sim$1 \\
Photon-recoil shift & $-$160$^\textrm{d}$ & & \\
\bottomrule
\end{tabular} }
{\small $^\textrm{a}$Dependent on the measurement. $^\textrm{b}$Averages out upon multiple realignments. $^\textrm{c}$For $v^+=0$ ($v^+=1$). $^\textrm{d}$Corresponds to $\tilde\nu_{\textrm{laser}}$ =12\,701.365 cm$^{-1}$.}
\end{table}

\begin{figure}

	\includegraphics[trim=0cm 0cm 0cm 0cm, clip=false, width=\linewidth]{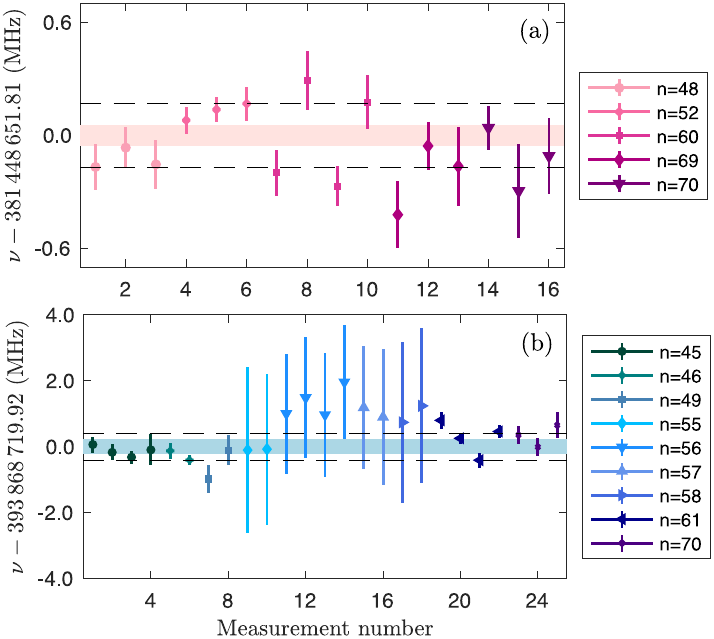}
        \caption{Fitted binding energies of the GK(0,2) [GK(2,2)] states of H$_2$ with respect to the $v^+=0$ (a)[$v^+=1$ (b)] ionization threshold. For each $n$, individual points correspond to measurements taken at different values of the electric field strength (in the range 90-300 mV/cm). The dashed lines indicate the weighted standard deviations of the data sets, and the shaded regions indicate the weighted standard deviations of the means (see text for detail).}
        \label{fig:resistark}
\end{figure}

\begin{table}
\centering
\caption{Summary of energy intervals used to determine the fundamental vibrational interval in H$_{2}^+$ and their uncertainties. $^\textrm{a}$The two uncertainties reported for the calculated value \cite{korobov17a} correspond to the theoretical uncertainty and the uncertainty in the proton-to-electron mass ratio.}
\label{tab:Evals}
\resizebox{3.4in}{!} {
\begin{tabular}{llll} 
\toprule
Energy interval & Value (cm$^{-1}$) & Ref. \\
\midrule
X$^+ (0,0) - $GK$(0,2) $ \  & 12\,723.757\,440\,7(18)$_\textrm{stat}$(40)$_\textrm{sys}$  & This work \\
X$^+ (0,0) - $GK$(0,2) $ \ & 12\,723.757\,461(23) & \cite{beyer19a} \\
 X$^+ (1,0) - $GK$(2,2) $ & 13\,138.046\,319(5)$_\textrm{stat}$(8)$_\textrm{sys}$  & This work \\
GK$(2,2) - $GK$(0,2) $ & \ \,\,1776.837\,736\,1(14) & \cite{hoelsch18a} \\
X$^+ (1,0) -$X$^+(0,0)$ &  \, \,2191.126\,614(5)$_\textrm{stat}$(12)$_\textrm{sys}$ & This work \\
X$^+ (1,0) -$X$^+(0,0)$ & \, \,2191.2(2)  & \cite{herzberg72a} \\
X$^+ (1,0) -$X$^+(0,0)$ & \, \,2191.126\,626\,344(17)(100)$^{\textrm{a}}$ & \cite{korobov17a} \\
\bottomrule
\end{tabular} }
\end{table}

The ionization energies for all investigated $n$ values (see color coding) and field strengths are displayed in Fig. \ref{fig:resistark}a) and b) for the $v^+=0$ and $v^+=1$ thresholds, respectively. The error budget corresponding to one such determined value of the ionization energy is presented in Table \ref{tab:errtr}. The statistical contributions to the uncertainties (vertical error bars) result from the least-squares fit described above. The uncertainties associated with the residual first-order Doppler shifts average out over the data set, which comprises the results of multiple measurements carried out after complete realignment of the optical layout, and are thus treated as statistical. In Fig. \ref{fig:resistark}, the dashed lines correspond to the weighted standard deviation of the data sets and the shaded regions to the weighted standard deviation of the mean, obtained by assuming that only measurements carried out after a full realignment of the experimental geometry are independent. Whereas the individual $v^+=0$ ionization energies have similar error bars and agree with each other within better than 400 kHz, several $v^+=1$ ionization energies, corresponding to $n$ = 55-58, exhibit large error bars and systematic shifts to higher values. At these $n$ values, the low-$\ell$ states are integrated in the Stark manifolds, leading to the intensity and lineshape perturbations described above. We attribute these discrepancies to the quantum defects and the exact wavefunctions of the perturbing low-$\ell$ states not being precisely known. The systematic deviations are compatible with the sensitivity analysis of the Stark-state positions on the quantum defects presented in Ref. \cite{hoelsch22a}. Because of the large error bars, these ionization energies hardly contribute to the weighted mean. The remaining systematic uncertainties, to which the lineshape model makes the largest contribution, are considered separately, as detailed in the lower part of Table \ref{tab:errtr}.

Table \ref{tab:Evals} compares the ionization energies of the GK(0,2) and GK(2,2) states (12\,723.757\,440\,7(18)$_\textrm{stat}$(40)$_\textrm{sys}$ cm$^{-1}$ and 13\,138.046\,319(5)$_\textrm{stat}$(8)$_\textrm{sys}$ cm$^{-1}$, respectively), as well as the fundamental vibrational wavenumber $\Delta G _{1/2} = 2191.126\,614(17)$ cm$^{-1}$ derived using Eq. \ref{eq:Enint}, with earlier values. The results obtained in the present work are all within about 1$\sigma$ of earlier theoretical \cite{korobov17a} and experimental \cite{beyer19a, herzberg72a} results. The fundamental vibrational interval determined here is four orders of magnitude more precise than the early pioneering experimental result (Ref. \cite{herzberg72a}), but still two orders of magnitude less precise than the theoretical value \cite{korobov17a}.

We have demonstrated the use of the ZQD method introduced in Ref. \cite{hoelsch22a} to determine, for the first time, the fundamental vibrational interval of H$_2^+$ at sub-MHz accuracy. The accuracy is currently limited by the uncertainties in the quantum defects of the $n$p and $n$d Rydberg series converging on the $v^+=1$ level of H$_2^+$. We expect that one to two orders of magnitude in precision could be gained by restricting the three-photon excitation to $m_{\ell} =3$ Rydberg-Stark states using circularly polarized radiation \cite{hogan09a}. Such states do not contain $\ell \leq 2$ character and their vibrational autoionization is suppressed. The method presented here is generally applicable. Because it does not rely on the existence of a permanent electric dipole moment, it is ideal for determining the rovibrational structures of homonuclear diatomic cations.

\section*{Acknowledgments}
We thank J. A. Agner and H. Schmutz for their contributions to setting up and maintaining the experimental infrastructure and Prof. Ch. Jungen for fruitful discussions. This work is supported financially by the Swiss National Science Foundation (grant no. 200020B-200478).

\newpage

\section{Supplementary Material}
This supplementary material provides information on the determination of the spectral positions of $n$p and $n$d Rydberg states of H$_2$ with a $v^+=0,1$ H$_2^+$ ion core.

\begin{figure*}[h]
  \begin{minipage}{0.28\textwidth}
    \centering
		\subfigure{\includegraphics[trim=0.1cm 1.45cm 0.4cm 0cm, clip=true, width=\linewidth]{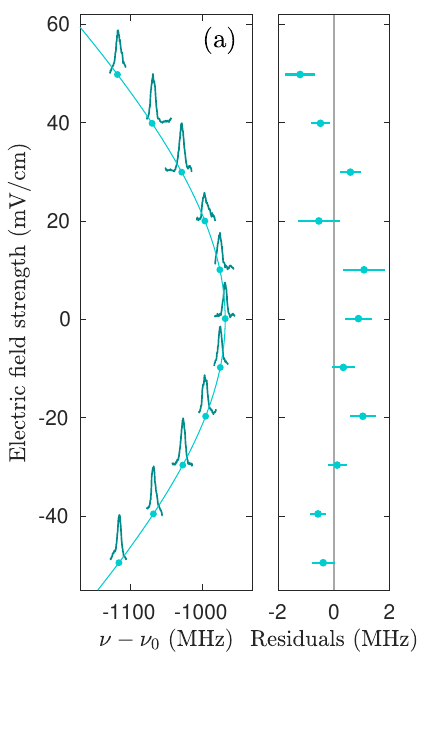}}
  \end{minipage}%
  \begin{minipage}{0.28\textwidth}
    \centering
		\subfigure{\includegraphics[trim=0.2cm 1.45cm 0.3cm 0cm, clip=true, width=\linewidth]{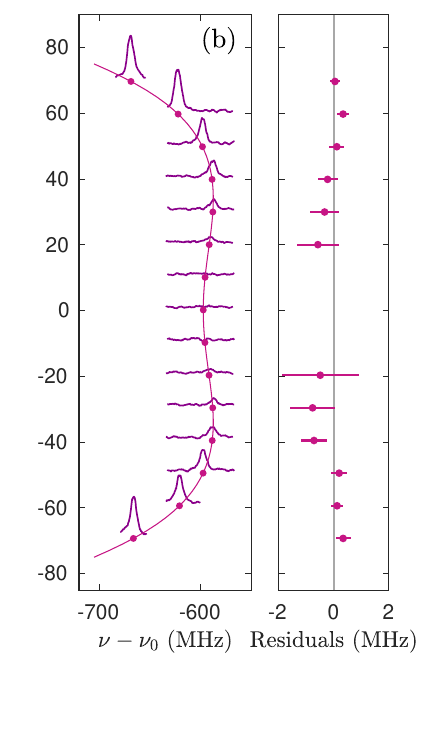}}
  \end{minipage}
      \caption{Spectra of the 70p0$_1$ ($v^+$=0) $\leftarrow$ GK(0,2) (a) and 70d0$_2$ ($v^+$=0) $\leftarrow$ GK(0,2) (b) transitions recorded in the presence of weak electric fields indicated along the vertical axes. Left panels: the dots indicate the line positions obtained in least-squares fits. The calculated Stark shifts are shown by the solid lines. Right panels: fit residuals.}
    \label{fig:spectrapd}
\end{figure*}

Accurate knowledge of the zero-field energies of $\ell \leq 2$ states is essential for the calculation of the Stark effect in high-$n$ Rydberg states of H$_2$
(see Ref. \cite{hoelsch22a} for details). The s Rydberg states are located so far from the high-$\ell$ manifolds that variations in their zero-field energies have a negligible effect on the calculation. The positions of most $n$p ($v^+=0$) Rydberg states are well known from previous studies \cite{sprecher14x}. To improve the accuracy of the calculation of the Stark effect, the positions of $n$p ($v^+=1$) and $n$d ($v^+=0,1$) Rydberg states were measured as illustrated in Fig. \ref{fig:spectrapd}. The figure displays the spectra of the 70p0$_1$ ($v^+$=0) $\leftarrow$ GK(0,2) (a) and 70d0$_2$ ($v^+$=0) $\leftarrow$ GK(0,2) (b) transitions, recorded in the presence of weak electric fields, and shows the extrapolation of the zero-field transition frequencies. The field-induced interaction of the 70d0$_2$ ($v^+$=0) state with the 70p0$_1$ ($v^+$=0) and 70f0$_3$ ($v^+$=0) states causes a field dependence which cannot be described by a quadratic function.

Table \ref{tab:Elowl} summarizes the binding energies of the p and d Rydberg states determined experimentally according to the procedure outlined above, and used in the calculations of the Stark effect. 

\begin{table}[h]
\caption{Binding energies $E_{\text{bind}}^ {v+}$ of  $n \ell 0 _{N=\ell} (v^+ = 0,1) $ Rydberg states of H$_2$ with respect to their respective $v^+$ ionization thresholds determined by extrapolation of their Stark shifts to zero field. The last column gives the respective uncertainties (1$\sigma$).}
\label{tab:Elowl}
\resizebox{3.25in}{!} {
\begin{tabular}{lll} 
\toprule
 $n \ell 0 _{N} (v^+) $ & \ \ \ \ \  \  $E_{\text{bind}}^ {v+} / h$ (MHz) & \ \ \  $\sigma _ {E_{\text{bind}}^ {v+}}$ (MHz)\\
 \midrule
52d0$_2$ ($v^+=0$) & \ \ \ \ \ \  1\,217\,707.5 & \ \ \   1.2 \\
60d0$_2$ ($v^+=0$) &  \ \ \ \ \ \ \ \ 914\,214.2 &  \ \ \   1.2\\
69d0$_2$ ($v^+=0$) &  \ \ \ \ \ \ \ \ 691\,421.8 &   \ \ \ 0.5 \\
70p0$_1$ ($v^+=0$) &   \ \ \ \ \ \ \ \  672\,181.5 &  \ \ \  0.6 \\
70d0$_2$ ($v^+=0$) &  \ \ \ \ \ \ \ \  671\,811.4 &  \ \ \ 0.5 \\
56d0$_2$ ($v^+=1$) &  \ \ \ \ \ \ 1\,049\,000 &  \ \ \  8 \\
61d0$_2$ ($v^+=1$) & \ \ \ \ \ \ \ \  884\,416 &  \ \ \ 9  \\
70p0$_1$ ($v^+=1$) & \ \ \ \ \ \ \ \  670\,241  &  \ \ \ 8 \\
70d0$_2$ ($v^+=1$) &  \ \ \ \ \ \ \ \ 671\,625 & \ \ \  6 \\
\bottomrule
\end{tabular} }
\end{table} 



\begin{thebibliography}{38}%
	\makeatletter
	\providecommand \@ifxundefined [1]{%
		\@ifx{#1\undefined}
	}%
	\providecommand \@ifnum [1]{%
		\ifnum #1\expandafter \@firstoftwo
		\else \expandafter \@secondoftwo
		\fi
	}%
	\providecommand \@ifx [1]{%
		\ifx #1\expandafter \@firstoftwo
		\else \expandafter \@secondoftwo
		\fi
	}%
	\providecommand \natexlab [1]{#1}%
	\providecommand \enquote  [1]{``#1''}%
	\providecommand \bibnamefont  [1]{#1}%
	\providecommand \bibfnamefont [1]{#1}%
	\providecommand \citenamefont [1]{#1}%
	\providecommand \href@noop [0]{\@secondoftwo}%
	\providecommand \href [0]{\begingroup \@sanitize@url \@href}%
	\providecommand \@href[1]{\@@startlink{#1}\@@href}%
	\providecommand \@@href[1]{\endgroup#1\@@endlink}%
	\providecommand \@sanitize@url [0]{\catcode `\\12\catcode `\$12\catcode
		`\&12\catcode `\#12\catcode `\^12\catcode `\_12\catcode `\%12\relax}%
	\providecommand \@@startlink[1]{}%
	\providecommand \@@endlink[0]{}%
	\providecommand \url  [0]{\begingroup\@sanitize@url \@url }%
	\providecommand \@url [1]{\endgroup\@href {#1}{\urlprefix }}%
	\providecommand \urlprefix  [0]{URL }%
	\providecommand \Eprint [0]{\href }%
	\providecommand \doibase [0]{https://doi.org/}%
	\providecommand \selectlanguage [0]{\@gobble}%
	\providecommand \bibinfo  [0]{\@secondoftwo}%
	\providecommand \bibfield  [0]{\@secondoftwo}%
	\providecommand \translation [1]{[#1]}%
	\providecommand \BibitemOpen [0]{}%
	\providecommand \bibitemStop [0]{}%
	\providecommand \bibitemNoStop [0]{.\EOS\space}%
	\providecommand \EOS [0]{\spacefactor3000\relax}%
	\providecommand \BibitemShut  [1]{\csname bibitem#1\endcsname}%
	\let\auto@bib@innerbib\@empty
	\bibitem [{\citenamefont {Bishop}\ and\ \citenamefont
		{Cheung}(1977)}]{bishop77a}%
	\BibitemOpen
	\bibfield  {author} {\bibinfo {author} {\bibfnamefont {D.~M.}\ \bibnamefont
			{Bishop}}\ and\ \bibinfo {author} {\bibfnamefont {L.~M.}\ \bibnamefont
			{Cheung}},\ }\bibfield  {title} {\bibinfo {title} {Calculation of transition
			frequencies for {H}$_{2}^{+}$ and its isotopes to spectroscopic accuracy},\
	}\href {https://doi.org/10.1103/PhysRevA.16.640} {\bibfield  {journal}
		{\bibinfo  {journal} {Phys. Rev. A}\ }\textbf {\bibinfo {volume} {16}},\
		\bibinfo {pages} {640} (\bibinfo {year} {1977})}\BibitemShut {NoStop}%
	\bibitem [{\citenamefont {Moss}(1993)}]{moss93a}%
	\BibitemOpen
	\bibfield  {author} {\bibinfo {author} {\bibfnamefont {R.~E.}\ \bibnamefont
			{Moss}},\ }\bibfield  {title} {\bibinfo {title} {Calculations for the
			vibration-rotation levels of {H$_2^+$} in its ground and first excited
			electronic states},\ }\href {https://doi.org/10.1080/00268979300103211}
	{\bibfield  {journal} {\bibinfo  {journal} {Mol. Phys.}\ }\textbf {\bibinfo
			{volume} {80}},\ \bibinfo {pages} {1541} (\bibinfo {year}
		{1993})}\BibitemShut {NoStop}%
	\bibitem [{\citenamefont {Korobov}(2006)}]{korobov06a}%
	\BibitemOpen
	\bibfield  {author} {\bibinfo {author} {\bibfnamefont {V.~I.}\ \bibnamefont
			{Korobov}},\ }\bibfield  {title} {\bibinfo {title} {Leading-order
			relativistic and radiative corrections to the rovibrational spectrum of
			{H$_2^+$} and {HD$^+$} molecular ions},\ }\href
	{https://doi.org/10.1103/PhysRevA.74.052506} {\bibfield  {journal} {\bibinfo
			{journal} {Phys. Rev. A}\ }\textbf {\bibinfo {volume} {74}},\ \bibinfo
		{pages} {052506} (\bibinfo {year} {2006})}\BibitemShut {NoStop}%
	\bibitem [{\citenamefont {Korobov}(2008)}]{korobov08a}%
	\BibitemOpen
	\bibfield  {author} {\bibinfo {author} {\bibfnamefont {V.~I.}\ \bibnamefont
			{Korobov}},\ }\bibfield  {title} {\bibinfo {title} {Relativistic corrections
			of {$m\alpha^6$} order to the rovibrational spectrum of {H$_2^+$} and
			{HD$^+$} molecular ions},\ }\href
	{https://doi.org/10.1103/PhysRevA.77.022509} {\bibfield  {journal} {\bibinfo
			{journal} {Phys. Rev. A}\ }\textbf {\bibinfo {volume} {77}},\ \bibinfo
		{pages} {022509} (\bibinfo {year} {2008})}\BibitemShut {NoStop}%
	\bibitem [{\citenamefont {Korobov}\ \emph {et~al.}(2017)\citenamefont
		{Korobov}, \citenamefont {Hilico},\ and\ \citenamefont {Karr}}]{korobov17a}%
	\BibitemOpen
	\bibfield  {author} {\bibinfo {author} {\bibfnamefont {V.~I.}\ \bibnamefont
			{Korobov}}, \bibinfo {author} {\bibfnamefont {L.}~\bibnamefont {Hilico}},\
		and\ \bibinfo {author} {\bibfnamefont {J.-{\mbox{Ph}}.}\ \bibnamefont
			{Karr}},\ }\bibfield  {title} {\bibinfo {title} {Fundamental transitions and
			ionization energies of the hydrogen molecular ions with few {ppt}
			uncertainty},\ }\href {https://doi.org/10.1103/PhysRevLett.118.233001}
	{\bibfield  {journal} {\bibinfo  {journal} {Phys. Rev. Lett.}\ }\textbf
		{\bibinfo {volume} {118}},\ \bibinfo {pages} {233001} (\bibinfo {year}
		{2017})}\BibitemShut {NoStop}%
	\bibitem [{\citenamefont {Mohr}\ \emph {et~al.}(2016)\citenamefont {Mohr},
		\citenamefont {Newell},\ and\ \citenamefont {Taylor}}]{mohr16a}%
	\BibitemOpen
	\bibfield  {author} {\bibinfo {author} {\bibfnamefont {P.~J.}\ \bibnamefont
			{Mohr}}, \bibinfo {author} {\bibfnamefont {D.~B.}\ \bibnamefont {Newell}},\
		and\ \bibinfo {author} {\bibfnamefont {B.~N.}\ \bibnamefont {Taylor}},\
	}\bibfield  {title} {\bibinfo {title} {{CODATA} recommended values of the
			fundamental physical constants: 2014},\ }\href
	{https://doi.org/10.1063/1.4954402} {\bibfield  {journal} {\bibinfo
			{journal} {J. Phys. Chem. Ref. Data}\ }\textbf {\bibinfo {volume} {45}},\
		\bibinfo {pages} {043102} (\bibinfo {year} {2016})}\BibitemShut {NoStop}%
	\bibitem [{\citenamefont {Herzberg}\ and\ \citenamefont
		{Jungen}(1972)}]{herzberg72a}%
	\BibitemOpen
	\bibfield  {author} {\bibinfo {author} {\bibfnamefont {G.}~\bibnamefont
			{Herzberg}}\ and\ \bibinfo {author} {\bibfnamefont
			{{\mbox{Ch}}.}~\bibnamefont {Jungen}},\ }\bibfield  {title} {\bibinfo {title}
		{{R}ydberg series and ionization potential of the {H$_2$} molecule},\ }\href
	{https://doi.org/10.1016/0022-2852(72)90064-1} {\bibfield  {journal}
		{\bibinfo  {journal} {J. Mol. Spectrosc.}\ }\textbf {\bibinfo {volume}
			{41}},\ \bibinfo {pages} {425} (\bibinfo {year} {1972})}\BibitemShut
	{NoStop}%
	\bibitem [{\citenamefont {Beyer}(2018)}]{beyer18e}%
	\BibitemOpen
	\bibfield  {author} {\bibinfo {author} {\bibfnamefont {M.}~\bibnamefont
			{Beyer}},\ }\emph {\bibinfo {title} {Precision spectroscopy and dynamics of
			molecular hydrogen and its ion}},\ \href
	{https://doi.org/10.3929/ethz-b-000307988} {Ph.D. thesis},\ \bibinfo
	{school} {Eidgen{\"{o}}ssische Technische Hochschule}, \bibinfo {address}
	{Z{\"{u}}rich, Switzerland} (\bibinfo {year} {2018}),\ \bibinfo {note}
	{{Diss. ETH Nr. 25199}}\BibitemShut {NoStop}%
	\bibitem [{\citenamefont {Karr}\ \emph {et~al.}(2016)\citenamefont {Karr},
		\citenamefont {Hilico}, \citenamefont {Koelemeij},\ and\ \citenamefont
		{Korobov}}]{karr16a}%
	\BibitemOpen
	\bibfield  {author} {\bibinfo {author} {\bibfnamefont {J.-{\mbox{Ph}}.}\
			\bibnamefont {Karr}}, \bibinfo {author} {\bibfnamefont {L.}~\bibnamefont
			{Hilico}}, \bibinfo {author} {\bibfnamefont {J.~C.~J.}\ \bibnamefont
			{Koelemeij}},\ and\ \bibinfo {author} {\bibfnamefont {V.~I.}\ \bibnamefont
			{Korobov}},\ }\bibfield  {title} {\bibinfo {title} {Hydrogen molecular ions
			for improved determination of fundamental constants},\ }\href
	{https://doi.org/10.1103/PhysRevA.94.050501} {\bibfield  {journal} {\bibinfo
			{journal} {Phys. Rev. A}\ }\textbf {\bibinfo {volume} {94}},\ \bibinfo
		{pages} {050501(R)} (\bibinfo {year} {2016})}\BibitemShut {NoStop}%
	\bibitem [{\citenamefont {Schmidt}\ \emph {et~al.}(2020)\citenamefont
		{Schmidt}, \citenamefont {Louvradoux}, \citenamefont {Heinrich},
		\citenamefont {Sillitoe}, \citenamefont {Simpson}, \citenamefont {Karr},\
		and\ \citenamefont {Hilico}}]{schmidt20a}%
	\BibitemOpen
	\bibfield  {author} {\bibinfo {author} {\bibfnamefont {J.}~\bibnamefont
			{Schmidt}}, \bibinfo {author} {\bibfnamefont {T.}~\bibnamefont {Louvradoux}},
		\bibinfo {author} {\bibfnamefont {J.}~\bibnamefont {Heinrich}}, \bibinfo
		{author} {\bibfnamefont {N.}~\bibnamefont {Sillitoe}}, \bibinfo {author}
		{\bibfnamefont {M.}~\bibnamefont {Simpson}}, \bibinfo {author} {\bibfnamefont
			{J.-P.}\ \bibnamefont {Karr}},\ and\ \bibinfo {author} {\bibfnamefont
			{L.}~\bibnamefont {Hilico}},\ }\bibfield  {title} {\bibinfo {title}
		{Trapping, cooling, and photodissociation analysis of state-selected
			{H}$_2^+$ ions produced by (3+1) multiphoton ionization},\ }\href
	{https://doi.org/10.1103/PhysRevApplied.14.024053} {\bibfield  {journal}
		{\bibinfo  {journal} {Phys. Rev. Appl.}\ }\textbf {\bibinfo {volume} {14}},\
		\bibinfo {pages} {024053} (\bibinfo {year} {2020})}\BibitemShut {NoStop}%
	\bibitem [{\citenamefont {Schwegler}\ \emph {et~al.}(2023)\citenamefont
		{Schwegler}, \citenamefont {Holzapfel}, \citenamefont {Stadler},
		\citenamefont {Mitjans}, \citenamefont {Sergachev}, \citenamefont {Home},\
		and\ \citenamefont {Kienzler}}]{schwegler23a}%
	\BibitemOpen
	\bibfield  {author} {\bibinfo {author} {\bibfnamefont {N.}~\bibnamefont
			{Schwegler}}, \bibinfo {author} {\bibfnamefont {D.}~\bibnamefont
			{Holzapfel}}, \bibinfo {author} {\bibfnamefont {M.}~\bibnamefont {Stadler}},
		\bibinfo {author} {\bibfnamefont {A.}~\bibnamefont {Mitjans}}, \bibinfo
		{author} {\bibfnamefont {I.}~\bibnamefont {Sergachev}}, \bibinfo {author}
		{\bibfnamefont {J.~P.}\ \bibnamefont {Home}},\ and\ \bibinfo {author}
		{\bibfnamefont {D.}~\bibnamefont {Kienzler}},\ }\bibfield  {title} {\bibinfo
		{title} {Trapping and ground-state cooling of a single
			${\mathrm{h}}_{2}^{+}$},\ }\href
	{https://doi.org/10.1103/PhysRevLett.131.133003} {\bibfield  {journal}
		{\bibinfo  {journal} {Phys. Rev. Lett.}\ }\textbf {\bibinfo {volume} {131}},\
		\bibinfo {pages} {133003} (\bibinfo {year} {2023})}\BibitemShut {NoStop}%
	\bibitem [{\citenamefont {Alighanbari}\ \emph {et~al.}(2023)\citenamefont
		{Alighanbari}, \citenamefont {Schenkel},\ and\ \citenamefont
		{Schiller}}]{alighanbari23a}%
	\BibitemOpen
	\bibfield  {author} {\bibinfo {author} {\bibfnamefont {S.}~\bibnamefont
			{Alighanbari}}, \bibinfo {author} {\bibfnamefont {M.}~\bibnamefont
			{Schenkel}},\ and\ \bibinfo {author} {\bibfnamefont {S.}~\bibnamefont
			{Schiller}},\ }\bibfield  {title} {\bibinfo {title} {{F}irst laser
			spectroscopy of a rovibrational transition in the molecular hydrogen ion
			${H}_2^+$},\ }in\ \href
	{https://konferenz.uni-hannover.de/event/79/book-of-abstracts.pdf} {\emph
		{\bibinfo {booktitle} {{Abstract, 7th European Conference on Trapped
					Ions}}}}\ (\bibinfo {year} {2023})\BibitemShut {NoStop}%
	\bibitem [{\citenamefont {Patra}\ \emph {et~al.}(2020)\citenamefont {Patra},
		\citenamefont {Germann}, \citenamefont {Karr}, \citenamefont {Haidar},
		\citenamefont {Hilico}, \citenamefont {Korobov}, \citenamefont {Cozijn},
		\citenamefont {Eikema}, \citenamefont {Ubachs},\ and\ \citenamefont
		{Koelemeij}}]{patra20a}%
	\BibitemOpen
	\bibfield  {author} {\bibinfo {author} {\bibfnamefont {S.}~\bibnamefont
			{Patra}}, \bibinfo {author} {\bibfnamefont {M.}~\bibnamefont {Germann}},
		\bibinfo {author} {\bibfnamefont {J.-P.}\ \bibnamefont {Karr}}, \bibinfo
		{author} {\bibfnamefont {M.}~\bibnamefont {Haidar}}, \bibinfo {author}
		{\bibfnamefont {L.}~\bibnamefont {Hilico}}, \bibinfo {author} {\bibfnamefont
			{V.~I.}\ \bibnamefont {Korobov}}, \bibinfo {author} {\bibfnamefont
			{F.~M.~J.}\ \bibnamefont {Cozijn}}, \bibinfo {author} {\bibfnamefont
			{K.~S.~E.}\ \bibnamefont {Eikema}}, \bibinfo {author} {\bibfnamefont
			{W.}~\bibnamefont {Ubachs}},\ and\ \bibinfo {author} {\bibfnamefont
			{J.~C.~J.}\ \bibnamefont {Koelemeij}},\ }\bibfield  {title} {\bibinfo {title}
		{Proton-electron mass ratio from laser spectroscopy of {HD$^+$} at the
			part-per-trillion level},\ }\href {https://doi.org/10.1126/science.aba0453}
	{\bibfield  {journal} {\bibinfo  {journal} {Science}\ }\textbf {\bibinfo
			{volume} {369}},\ \bibinfo {pages} {1238} (\bibinfo {year}
		{2020})}\BibitemShut {NoStop}%
	\bibitem [{\citenamefont {Alighanbari}\ \emph {et~al.}(2020)\citenamefont
		{Alighanbari}, \citenamefont {Giri}, \citenamefont {Constantin},
		\citenamefont {Korobov},\ and\ \citenamefont {Schiller}}]{alighanbari20a}%
	\BibitemOpen
	\bibfield  {author} {\bibinfo {author} {\bibfnamefont {S.}~\bibnamefont
			{Alighanbari}}, \bibinfo {author} {\bibfnamefont {G.~S.}\ \bibnamefont
			{Giri}}, \bibinfo {author} {\bibfnamefont {F.~L.}\ \bibnamefont
			{Constantin}}, \bibinfo {author} {\bibfnamefont {V.~I.}\ \bibnamefont
			{Korobov}},\ and\ \bibinfo {author} {\bibfnamefont {S.}~\bibnamefont
			{Schiller}},\ }\bibfield  {title} {\bibinfo {title} {Precise test of quantum
			electrodynamics and determination of fundamental constants with {HD$^+$}
			ions},\ }\href {https://doi.org/10.1038/s41586-020-2261-5} {\bibfield
		{journal} {\bibinfo  {journal} {Nature}\ }\textbf {\bibinfo {volume} {581}},\
		\bibinfo {pages} {152} (\bibinfo {year} {2020})}\BibitemShut {NoStop}%
	\bibitem [{\citenamefont {Kortunov}\ \emph {et~al.}(2021)\citenamefont
		{Kortunov}, \citenamefont {Alighanbari}, \citenamefont {Hansen},
		\citenamefont {Giri}, \citenamefont {Korobov},\ and\ \citenamefont
		{Schiller}}]{kortunov21a}%
	\BibitemOpen
	\bibfield  {author} {\bibinfo {author} {\bibfnamefont {I.~V.}\ \bibnamefont
			{Kortunov}}, \bibinfo {author} {\bibfnamefont {S.}~\bibnamefont
			{Alighanbari}}, \bibinfo {author} {\bibfnamefont {M.~G.}\ \bibnamefont
			{Hansen}}, \bibinfo {author} {\bibfnamefont {G.~S.}\ \bibnamefont {Giri}},
		\bibinfo {author} {\bibfnamefont {V.~I.}\ \bibnamefont {Korobov}},\ and\
		\bibinfo {author} {\bibfnamefont {S.}~\bibnamefont {Schiller}},\ }\bibfield
	{title} {\bibinfo {title} {Proton--electron mass ratio by high-resolution
			optical spectroscopy of ion ensembles in the resolved-carrier regime},\
	}\href {https://doi.org/10.1038/s41567-020-01150-7} {\bibfield  {journal}
		{\bibinfo  {journal} {Nat. Phys.}\ }\textbf {\bibinfo {volume} {17}},\
		\bibinfo {pages} {569} (\bibinfo {year} {2021})}\BibitemShut {NoStop}%
	\bibitem [{\citenamefont {Germann}\ \emph {et~al.}(2021)\citenamefont
		{Germann}, \citenamefont {Patra}, \citenamefont {Karr}, \citenamefont
		{Hilico}, \citenamefont {Korobov}, \citenamefont {Salumbides}, \citenamefont
		{Eikema}, \citenamefont {Ubachs},\ and\ \citenamefont
		{Koelemeij}}]{germann21a}%
	\BibitemOpen
	\bibfield  {author} {\bibinfo {author} {\bibfnamefont {M.}~\bibnamefont
			{Germann}}, \bibinfo {author} {\bibfnamefont {S.}~\bibnamefont {Patra}},
		\bibinfo {author} {\bibfnamefont {J.-{\mbox{Ph}}.}\ \bibnamefont {Karr}},
		\bibinfo {author} {\bibfnamefont {L.}~\bibnamefont {Hilico}}, \bibinfo
		{author} {\bibfnamefont {V.~I.}\ \bibnamefont {Korobov}}, \bibinfo {author}
		{\bibfnamefont {E.~J.}\ \bibnamefont {Salumbides}}, \bibinfo {author}
		{\bibfnamefont {K.~S.~E.}\ \bibnamefont {Eikema}}, \bibinfo {author}
		{\bibfnamefont {W.}~\bibnamefont {Ubachs}},\ and\ \bibinfo {author}
		{\bibfnamefont {J.~C.~J.}\ \bibnamefont {Koelemeij}},\ }\bibfield  {title}
	{\bibinfo {title} {Three-body {QED} test and fifth-force constraint from
			vibrations and rotations of {HD}$^+$},\ }\href
	{https://doi.org/10.1103/PhysRevResearch.3.L022028} {\bibfield  {journal}
		{\bibinfo  {journal} {Phys. Rev. Research}\ }\textbf {\bibinfo {volume}
			{3}},\ \bibinfo {pages} {L022028} (\bibinfo {year} {2021})}\BibitemShut
	{NoStop}%
	\bibitem [{\citenamefont {Koelemeij}(2022)}]{koelemeij22a}%
	\BibitemOpen
	\bibfield  {author} {\bibinfo {author} {\bibfnamefont {J.~C.~J.}\
			\bibnamefont {Koelemeij}},\ }\bibfield  {title} {\bibinfo {title} {Effect of
			correlated hyperfine theory errors in the determination of rotational and
			vibrational transition frequencies in {HD}$^+$},\ }\href
	{https://doi.org/10.1080/00268976.2022.2058637} {\bibfield  {journal}
		{\bibinfo  {journal} {Mol. Phys.}\ }\textbf {\bibinfo {volume} {120}},\
		\bibinfo {pages} {e2058637} (\bibinfo {year} {2022})}\BibitemShut {NoStop}%
	\bibitem [{\citenamefont {Karr}\ and\ \citenamefont
		{Koelemeij}(2023)}]{karr23a}%
	\BibitemOpen
	\bibfield  {author} {\bibinfo {author} {\bibfnamefont {J.-P.}\ \bibnamefont
			{Karr}}\ and\ \bibinfo {author} {\bibfnamefont {J.~C.~J.}\ \bibnamefont
			{Koelemeij}},\ }\bibfield  {title} {\bibinfo {title} {Extraction of
			spin-averaged rovibrational transition frequencies in {HD}$^+$ for the
			determination of fundamental constants},\ }\href
	{https://doi.org/10.1080/00268976.2023.2216081} {\bibfield  {journal}
		{\bibinfo  {journal} {Mol. Phys.}\ }\textbf {\bibinfo {volume} {121}},\
		\bibinfo {pages} {e2216081} (\bibinfo {year} {2023})}\BibitemShut {NoStop}%
	\bibitem [{\citenamefont {H{\"{o}}lsch}\ \emph {et~al.}(2022)\citenamefont
		{H{\"{o}}lsch}, \citenamefont {Doran}, \citenamefont {Beyer},\ and\
		\citenamefont {Merkt}}]{hoelsch22a}%
	\BibitemOpen
	\bibfield  {author} {\bibinfo {author} {\bibfnamefont {N.}~\bibnamefont
			{H{\"{o}}lsch}}, \bibinfo {author} {\bibfnamefont {I.}~\bibnamefont {Doran}},
		\bibinfo {author} {\bibfnamefont {M.}~\bibnamefont {Beyer}},\ and\ \bibinfo
		{author} {\bibfnamefont {F.}~\bibnamefont {Merkt}},\ }\bibfield  {title}
	{\bibinfo {title} {Precision millimetre-wave spectroscopy and calculation of
			the {S}tark manifolds in high {R}ydberg states of para-{H$_2$}},\ }\href
	{https://doi.org/10.1016/j.jms.2022.111648} {\bibfield  {journal} {\bibinfo
			{journal} {J. Mol. Spectrosc.}\ }\textbf {\bibinfo {volume} {387}},\ \bibinfo
		{pages} {111648} (\bibinfo {year} {2022})}\BibitemShut {NoStop}%
	\bibitem [{\citenamefont {Jungen}(2011)}]{jungen11a}%
	\BibitemOpen
	\bibfield  {author} {\bibinfo {author} {\bibfnamefont
			{{\mbox{Ch}}.}~\bibnamefont {Jungen}},\ }\bibfield  {title} {\bibinfo {title}
		{Elements of quantum defect theory},\ }in\ \href
	{https://doi.org/10.1002/9780470749593.hrs024} {\emph {\bibinfo {booktitle}
			{Handbook of High-Resolution Spectroscopy}}},\ Vol.~\bibinfo {volume} {1},\
	\bibinfo {editor} {edited by\ \bibinfo {editor} {\bibfnamefont
			{M.}~\bibnamefont {Quack}}\ and\ \bibinfo {editor} {\bibfnamefont
			{F.}~\bibnamefont {Merkt}}}\ (\bibinfo  {publisher} {John Wiley \& Sons},\
	\bibinfo {address} {Chichester},\ \bibinfo {year} {2011})\ pp.\ \bibinfo
	{pages} {471--510}\BibitemShut {NoStop}%
	\bibitem [{\citenamefont {Osterwalder}\ \emph {et~al.}(2004)\citenamefont
		{Osterwalder}, \citenamefont {W{\"{u}}est}, \citenamefont {Merkt},\ and\
		\citenamefont {Jungen}}]{osterwalder04a}%
	\BibitemOpen
	\bibfield  {author} {\bibinfo {author} {\bibfnamefont {A.}~\bibnamefont
			{Osterwalder}}, \bibinfo {author} {\bibfnamefont {A.}~\bibnamefont
			{W{\"{u}}est}}, \bibinfo {author} {\bibfnamefont {F.}~\bibnamefont {Merkt}},\
		and\ \bibinfo {author} {\bibfnamefont {{\mbox{Ch}}.}~\bibnamefont {Jungen}},\
	}\bibfield  {title} {\bibinfo {title} {High-resolution millimeter wave
			spectroscopy and multichannel quantum defect theory of the hyperfine
			structure in high {R}ydberg states of molecular hydrogen {H$_2$}},\ }\href
	{https://doi.org/10.1063/1.1792596} {\bibfield  {journal} {\bibinfo
			{journal} {J. Chem. Phys.}\ }\textbf {\bibinfo {volume} {121}},\ \bibinfo
		{pages} {11810} (\bibinfo {year} {2004})}\BibitemShut {NoStop}%
	\bibitem [{\citenamefont {Sprecher}\ \emph {et~al.}(2014)\citenamefont
		{Sprecher}, \citenamefont {Liu}, \citenamefont {Kr{\"{a}}henmann},
		\citenamefont {Sch{\"{a}}fer},\ and\ \citenamefont {Merkt}}]{sprecher14a}%
	\BibitemOpen
	\bibfield  {author} {\bibinfo {author} {\bibfnamefont {D.}~\bibnamefont
			{Sprecher}}, \bibinfo {author} {\bibfnamefont {J.}~\bibnamefont {Liu}},
		\bibinfo {author} {\bibfnamefont {T.}~\bibnamefont {Kr{\"{a}}henmann}},
		\bibinfo {author} {\bibfnamefont {M.}~\bibnamefont {Sch{\"{a}}fer}},\ and\
		\bibinfo {author} {\bibfnamefont {F.}~\bibnamefont {Merkt}},\ }\bibfield
	{title} {\bibinfo {title} {High-resolution spectroscopy and quantum-defect
			model for the \textit{gerade} triplet {$np$} and {$nf$} {R}ydberg states of
			{He$_{2}$}},\ }\href {https://doi.org/10.1063/1.4864002} {\bibfield
		{journal} {\bibinfo  {journal} {J. Chem. Phys.}\ }\textbf {\bibinfo {volume}
			{140}},\ \bibinfo {pages} {064304} (\bibinfo {year} {2014})}\BibitemShut
	{NoStop}%
	\bibitem [{\citenamefont {Eyler}\ and\ \citenamefont
		{Pipkin}(1983)}]{eyler83a}%
	\BibitemOpen
	\bibfield  {author} {\bibinfo {author} {\bibfnamefont {E.~E.}\ \bibnamefont
			{Eyler}}\ and\ \bibinfo {author} {\bibfnamefont {F.~M.}\ \bibnamefont
			{Pipkin}},\ }\bibfield  {title} {\bibinfo {title} {Triplet {4$d$} states of
			{H$_2$}: Experimental observation and comparison with an \textit{ab initio}
			model for {R}ydberg-state energies},\ }\href
	{https://doi.org/10.1103/PhysRevA.27.2462} {\bibfield  {journal} {\bibinfo
			{journal} {Phys. Rev. A}\ }\textbf {\bibinfo {volume} {27}},\ \bibinfo
		{pages} {2462} (\bibinfo {year} {1983})}\BibitemShut {NoStop}%
	\bibitem [{\citenamefont {Beyer}\ \emph {et~al.}(2018)\citenamefont {Beyer},
		\citenamefont {H{\"{o}}lsch}, \citenamefont {Agner}, \citenamefont
		{Deiglmayr}, \citenamefont {Schmutz},\ and\ \citenamefont
		{Merkt}}]{beyer18a}%
	\BibitemOpen
	\bibfield  {author} {\bibinfo {author} {\bibfnamefont {M.}~\bibnamefont
			{Beyer}}, \bibinfo {author} {\bibfnamefont {N.}~\bibnamefont {H{\"{o}}lsch}},
		\bibinfo {author} {\bibfnamefont {J.~A.}\ \bibnamefont {Agner}}, \bibinfo
		{author} {\bibfnamefont {J.}~\bibnamefont {Deiglmayr}}, \bibinfo {author}
		{\bibfnamefont {H.}~\bibnamefont {Schmutz}},\ and\ \bibinfo {author}
		{\bibfnamefont {F.}~\bibnamefont {Merkt}},\ }\bibfield  {title} {\bibinfo
		{title} {Metrology of high-{$n$} {R}ydberg states of molecular hydrogen with
			{$\Delta\nu/\nu=2\times10^{-10}$} accuracy},\ }\href
	{https://doi.org/10.1103/PhysRevA.97.012501} {\bibfield  {journal} {\bibinfo
			{journal} {Phys. Rev. A}\ }\textbf {\bibinfo {volume} {97}},\ \bibinfo
		{pages} {012501} (\bibinfo {year} {2018})}\BibitemShut {NoStop}%
	\bibitem [{\citenamefont {H{\"{o}}lsch}\ \emph {et~al.}(2018)\citenamefont
		{H{\"{o}}lsch}, \citenamefont {Beyer},\ and\ \citenamefont
		{Merkt}}]{hoelsch18a}%
	\BibitemOpen
	\bibfield  {author} {\bibinfo {author} {\bibfnamefont {N.}~\bibnamefont
			{H{\"{o}}lsch}}, \bibinfo {author} {\bibfnamefont {M.}~\bibnamefont
			{Beyer}},\ and\ \bibinfo {author} {\bibfnamefont {F.}~\bibnamefont {Merkt}},\
	}\bibfield  {title} {\bibinfo {title} {Nonadiabatic effects on the positions
			and lifetimes of the low-lying rovibrational levels of the {GK
				$^1\Sigma_{\mathrm g}^+$} and {H $^1\Sigma_{\mathrm g}^+$} states of
			{H$_2$}},\ }\href {https://doi.org/10.1039/c8cp05233f} {\bibfield  {journal}
		{\bibinfo  {journal} {Phys. Chem. Chem. Phys.}\ }\textbf {\bibinfo {volume}
			{20}},\ \bibinfo {pages} {26837} (\bibinfo {year} {2018})}\BibitemShut
	{NoStop}%
	\bibitem [{\citenamefont {Bethe}\ and\ \citenamefont
		{Salpeter}(1957)}]{bethe57a}%
	\BibitemOpen
	\bibfield  {author} {\bibinfo {author} {\bibfnamefont {H.~A.}\ \bibnamefont
			{Bethe}}\ and\ \bibinfo {author} {\bibfnamefont {E.~E.}\ \bibnamefont
			{Salpeter}},\ }\href {https://doi.org/10.1007/978-1-4613-4104-8} {\emph
		{\bibinfo {title} {{Quantum Mechanics of One- and Two-Electron Atoms}}}}\
	(\bibinfo  {publisher} {Springer},\ \bibinfo {address} {Berlin},\ \bibinfo
	{year} {1957})\BibitemShut {NoStop}%
	\bibitem [{\citenamefont {Fano}(1961)}]{fano61a}%
	\BibitemOpen
	\bibfield  {author} {\bibinfo {author} {\bibfnamefont {U.}~\bibnamefont
			{Fano}},\ }\bibfield  {title} {\bibinfo {title} {Effects of configuration
			interaction on intensities and phase shifts},\ }\href
	{https://doi.org/10.1103/PhysRev.124.1866} {\bibfield  {journal} {\bibinfo
			{journal} {Phys. Rev.}\ }\textbf {\bibinfo {volume} {124}},\ \bibinfo {pages}
		{1866} (\bibinfo {year} {1961})}\BibitemShut {NoStop}%
	\bibitem [{\citenamefont {Jungen}\ and\ \citenamefont
		{Raoult}(1981)}]{jungen81a}%
	\BibitemOpen
	\bibfield  {author} {\bibinfo {author} {\bibfnamefont
			{{\mbox{Ch}}.}~\bibnamefont {Jungen}}\ and\ \bibinfo {author} {\bibfnamefont
			{M.}~\bibnamefont {Raoult}},\ }\bibfield  {title} {\bibinfo {title}
		{Spectroscopy in the ionisation continuum. vibrational preionisation in
			{H$_2$} calculated by multichannel quantum-defect theory},\ }\href
	{https://doi.org/10.1039/DC9817100253} {\bibfield  {journal} {\bibinfo
			{journal} {Faraday Discuss.}\ }\textbf {\bibinfo {volume} {71}},\ \bibinfo
		{pages} {253} (\bibinfo {year} {1981})}\BibitemShut {NoStop}%
	\bibitem [{\citenamefont {Gounand}\ \emph {et~al.}(1983)\citenamefont
		{Gounand}, \citenamefont {Gallagher}, \citenamefont {Sandner}, \citenamefont
		{Safinya},\ and\ \citenamefont {Kachru}}]{gounand83a}%
	\BibitemOpen
	\bibfield  {author} {\bibinfo {author} {\bibfnamefont {F.}~\bibnamefont
			{Gounand}}, \bibinfo {author} {\bibfnamefont {T.~F.}\ \bibnamefont
			{Gallagher}}, \bibinfo {author} {\bibfnamefont {W.}~\bibnamefont {Sandner}},
		\bibinfo {author} {\bibfnamefont {K.~A.}\ \bibnamefont {Safinya}},\ and\
		\bibinfo {author} {\bibfnamefont {R.}~\bibnamefont {Kachru}},\ }\bibfield
	{title} {\bibinfo {title} {Interaction between two {R}ydberg series of
			autoionizing levels in barium},\ }\href
	{https://doi.org/10.1103/PhysRevA.27.1925} {\bibfield  {journal} {\bibinfo
			{journal} {Phys. Rev. A}\ }\textbf {\bibinfo {volume} {27}},\ \bibinfo
		{pages} {1925} (\bibinfo {year} {1983})}\BibitemShut {NoStop}%
	\bibitem [{\citenamefont {Du}\ and\ \citenamefont {Greene}(1986)}]{du86a}%
	\BibitemOpen
	\bibfield  {author} {\bibinfo {author} {\bibfnamefont {N.~Y.}\ \bibnamefont
			{Du}}\ and\ \bibinfo {author} {\bibfnamefont {{\mbox{Ch}}.~H.}\ \bibnamefont
			{Greene}},\ }\bibfield  {title} {\bibinfo {title} {Quantum defect analysis of
			{HD} photoionization},\ }\href {https://doi.org/10.1063/1.451553} {\bibfield
		{journal} {\bibinfo  {journal} {J. Chem. Phys.}\ }\textbf {\bibinfo {volume}
			{85}},\ \bibinfo {pages} {5430} (\bibinfo {year} {1986})}\BibitemShut
	{NoStop}%
	\bibitem [{\citenamefont {Kung}\ \emph {et~al.}(1986)\citenamefont {Kung},
		\citenamefont {Page}, \citenamefont {Larkin}, \citenamefont {Shen},\ and\
		\citenamefont {Lee}}]{kung86a}%
	\BibitemOpen
	\bibfield  {author} {\bibinfo {author} {\bibfnamefont {A.~H.}\ \bibnamefont
			{Kung}}, \bibinfo {author} {\bibfnamefont {R.~H.}\ \bibnamefont {Page}},
		\bibinfo {author} {\bibfnamefont {R.~J.}\ \bibnamefont {Larkin}}, \bibinfo
		{author} {\bibfnamefont {Y.~R.}\ \bibnamefont {Shen}},\ and\ \bibinfo
		{author} {\bibfnamefont {Y.~T.}\ \bibnamefont {Lee}},\ }\bibfield  {title}
	{\bibinfo {title} {{R}ydberg spectroscopy of {H$_2$} via stepwise resonant
			two-photon ion-pair {(H$^+$+H$^-$)} production},\ }\href
	{https://doi.org/10.1103/PhysRevLett.56.328} {\bibfield  {journal} {\bibinfo
			{journal} {Phys. Rev. Lett.}\ }\textbf {\bibinfo {volume} {56}},\ \bibinfo
		{pages} {328} (\bibinfo {year} {1986})}\BibitemShut {NoStop}%
	\bibitem [{\citenamefont {Domke}\ \emph {et~al.}(1991)\citenamefont {Domke},
		\citenamefont {Xue}, \citenamefont {Puschmann}, \citenamefont {Mandel},
		\citenamefont {Hudson}, \citenamefont {Shirley}, \citenamefont {Kaindl},
		\citenamefont {Greene}, \citenamefont {Sadeghpour},\ and\ \citenamefont
		{Petersen}}]{domke91a}%
	\BibitemOpen
	\bibfield  {author} {\bibinfo {author} {\bibfnamefont {M.}~\bibnamefont
			{Domke}}, \bibinfo {author} {\bibfnamefont {C.}~\bibnamefont {Xue}}, \bibinfo
		{author} {\bibfnamefont {A.}~\bibnamefont {Puschmann}}, \bibinfo {author}
		{\bibfnamefont {T.}~\bibnamefont {Mandel}}, \bibinfo {author} {\bibfnamefont
			{E.}~\bibnamefont {Hudson}}, \bibinfo {author} {\bibfnamefont {D.~A.}\
			\bibnamefont {Shirley}}, \bibinfo {author} {\bibfnamefont {G.}~\bibnamefont
			{Kaindl}}, \bibinfo {author} {\bibfnamefont {C.~H.}\ \bibnamefont {Greene}},
		\bibinfo {author} {\bibfnamefont {H.~R.}\ \bibnamefont {Sadeghpour}},\ and\
		\bibinfo {author} {\bibfnamefont {H.}~\bibnamefont {Petersen}},\ }\bibfield
	{title} {\bibinfo {title} {Extensive double-excitation states in atomic
			helium},\ }\href {https://doi.org/10.1103/PhysRevLett.66.1306} {\bibfield
		{journal} {\bibinfo  {journal} {Phys. Rev. Lett.}\ }\textbf {\bibinfo
			{volume} {66}},\ \bibinfo {pages} {1306} (\bibinfo {year}
		{1991})}\BibitemShut {NoStop}%
	\bibitem [{\citenamefont {Kim}\ and\ \citenamefont {Yoshihara}(1993)}]{kim93a}%
	\BibitemOpen
	\bibfield  {author} {\bibinfo {author} {\bibfnamefont {B.}~\bibnamefont
			{Kim}}\ and\ \bibinfo {author} {\bibfnamefont {K.}~\bibnamefont
			{Yoshihara}},\ }\bibfield  {title} {\bibinfo {title} {{Multichannel quantum
				interference in the predissociation of Cs$_2$: Observation of q-reversal in a
				complex resonance}},\ }\href {https://doi.org/10.1063/1.465336} {\bibfield
		{journal} {\bibinfo  {journal} {J. Chem. Phys.}\ }\textbf {\bibinfo {volume}
			{99}},\ \bibinfo {pages} {1433} (\bibinfo {year} {1993})}\BibitemShut
	{NoStop}%
	\bibitem [{\citenamefont {Greetham}\ \emph {et~al.}(2003)\citenamefont
		{Greetham}, \citenamefont {Hollenstein}, \citenamefont {Seiler},
		\citenamefont {Ubachs},\ and\ \citenamefont {Merkt}}]{greetham03a}%
	\BibitemOpen
	\bibfield  {author} {\bibinfo {author} {\bibfnamefont {G.~M.}\ \bibnamefont
			{Greetham}}, \bibinfo {author} {\bibfnamefont {U.}~\bibnamefont
			{Hollenstein}}, \bibinfo {author} {\bibfnamefont {R.}~\bibnamefont {Seiler}},
		\bibinfo {author} {\bibfnamefont {W.}~\bibnamefont {Ubachs}},\ and\ \bibinfo
		{author} {\bibfnamefont {F.}~\bibnamefont {Merkt}},\ }\bibfield  {title}
	{\bibinfo {title} {High-resolution {VUV} photoionization spectroscopy of {HD}
			between the {X $^2\Sigma_{\mathrm{g}}^+$ $v^+=0$} and {$v^+=1$} thresholds},\
	}\href {https://doi.org/10.1039/b302876n} {\bibfield  {journal} {\bibinfo
			{journal} {Phys. Chem. Chem. Phys.}\ }\textbf {\bibinfo {volume} {5}},\
		\bibinfo {pages} {2528} (\bibinfo {year} {2003})}\BibitemShut {NoStop}%
	\bibitem [{\citenamefont {Viteri}\ \emph {et~al.}(2007)\citenamefont {Viteri},
		\citenamefont {Gilkison}, \citenamefont {Schr{\"{o}}der},\ and\ \citenamefont
		{Grant}}]{viteri07b}%
	\BibitemOpen
	\bibfield  {author} {\bibinfo {author} {\bibfnamefont {C.~R.}\ \bibnamefont
			{Viteri}}, \bibinfo {author} {\bibfnamefont {A.~T.}\ \bibnamefont
			{Gilkison}}, \bibinfo {author} {\bibfnamefont {F.~S.}\ \bibnamefont
			{Schr{\"{o}}der}},\ and\ \bibinfo {author} {\bibfnamefont {E.~R.}\
			\bibnamefont {Grant}},\ }\bibfield  {title} {\bibinfo {title}
		{Discrete-continuum and discrete-discrete interactions in the autoionization
			spectrum of $^{11}${BH}},\ }\href {https://doi.org/10.1080/00268970701361314}
	{\bibfield  {journal} {\bibinfo  {journal} {Mol. Phys.}\ }\textbf {\bibinfo
			{volume} {105}},\ \bibinfo {pages} {1589} (\bibinfo {year}
		{2007})}\BibitemShut {NoStop}%
	\bibitem [{\citenamefont {Wang}\ \emph {et~al.}(2017)\citenamefont {Wang},
		\citenamefont {Meng},\ and\ \citenamefont {Mo}}]{wang17a}%
	\BibitemOpen
	\bibfield  {author} {\bibinfo {author} {\bibfnamefont {J.}~\bibnamefont
			{Wang}}, \bibinfo {author} {\bibfnamefont {Q.}~\bibnamefont {Meng}},\ and\
		\bibinfo {author} {\bibfnamefont {Y.}~\bibnamefont {Mo}},\ }\bibfield
	{title} {\bibinfo {title} {{Electronic and tunneling predissociations in the
				2p$\pi$\,C\, $^1\Pi_u^\pm(v=19)$ and 3p$\pi$\,D\, $^1\Pi_u^\pm(v=4,5)$ states
				ofD$_2$ studied by a combination of XUV laser and velocity map imaging}},\
	}\href {https://doi.org/10.1021/acs.jpca.7b04808} {\bibfield  {journal}
		{\bibinfo  {journal} {J. Phys. Chem. A}\ }\textbf {\bibinfo {volume} {121}},\
		\bibinfo {pages} {5785} (\bibinfo {year} {2017})}\BibitemShut {NoStop}%
	\bibitem [{\citenamefont {Beyer}\ \emph {et~al.}(2019)\citenamefont {Beyer},
		\citenamefont {H{\"{o}}lsch}, \citenamefont {Hussels}, \citenamefont {Cheng},
		\citenamefont {Salumbides}, \citenamefont {Eikema}, \citenamefont {Ubachs},
		\citenamefont {Jungen},\ and\ \citenamefont {Merkt}}]{beyer19a}%
	\BibitemOpen
	\bibfield  {author} {\bibinfo {author} {\bibfnamefont {M.}~\bibnamefont
			{Beyer}}, \bibinfo {author} {\bibfnamefont {N.}~\bibnamefont {H{\"{o}}lsch}},
		\bibinfo {author} {\bibfnamefont {J.}~\bibnamefont {Hussels}}, \bibinfo
		{author} {\bibfnamefont {C.-F.}\ \bibnamefont {Cheng}}, \bibinfo {author}
		{\bibfnamefont {E.~J.}\ \bibnamefont {Salumbides}}, \bibinfo {author}
		{\bibfnamefont {K.~S.~E.}\ \bibnamefont {Eikema}}, \bibinfo {author}
		{\bibfnamefont {W.}~\bibnamefont {Ubachs}}, \bibinfo {author} {\bibfnamefont
			{{\mbox{Ch}}.}~\bibnamefont {Jungen}},\ and\ \bibinfo {author} {\bibfnamefont
			{F.}~\bibnamefont {Merkt}},\ }\bibfield  {title} {\bibinfo {title}
		{Determination of the interval between the ground states of para- and
			ortho-{H$_2$}},\ }\href {https://doi.org/10.1103/PhysRevLett.123.163002}
	{\bibfield  {journal} {\bibinfo  {journal} {Phys. Rev. Lett.}\ }\textbf
		{\bibinfo {volume} {123}},\ \bibinfo {pages} {163002} (\bibinfo {year}
		{2019})}\BibitemShut {NoStop}%
	\bibitem [{\citenamefont {Hogan}\ \emph {et~al.}(2009)\citenamefont {Hogan},
		\citenamefont {Seiler},\ and\ \citenamefont {Merkt}}]{hogan09a}%
	\BibitemOpen
	\bibfield  {author} {\bibinfo {author} {\bibfnamefont {S.~D.}\ \bibnamefont
			{Hogan}}, \bibinfo {author} {\bibfnamefont {{\mbox{Ch}}.}~\bibnamefont
			{Seiler}},\ and\ \bibinfo {author} {\bibfnamefont {F.}~\bibnamefont
			{Merkt}},\ }\bibfield  {title} {\bibinfo {title} {{R}ydberg-state-enabled
			deceleration and trapping of cold molecules},\ }\href
	{https://doi.org/10.1103/PhysRevLett.103.123001} {\bibfield  {journal}
		{\bibinfo  {journal} {Phys. Rev. Lett.}\ }\textbf {\bibinfo {volume} {103}},\
		\bibinfo {pages} {123001} (\bibinfo {year} {2009})}\BibitemShut {NoStop}%
		\bibitem [{\citenamefont {Sprecher}\ \emph {et~al.}(2014)\citenamefont
  {Sprecher}, \citenamefont {Jungen},\ and\ \citenamefont
  {Merkt}}]{sprecher14x}%
  \BibitemOpen
  \bibfield  {author} {\bibinfo {author} {\bibfnamefont {D.}~\bibnamefont
  {Sprecher}}, \bibinfo {author} {\bibfnamefont {{\mbox{Ch}}.}~\bibnamefont
  {Jungen}},\ and\ \bibinfo {author} {\bibfnamefont {F.}~\bibnamefont
  {Merkt}},\ }\bibfield  {title} {\bibinfo {title} {Determination of the
  binding energies of the {$np$} {R}ydberg states of {H$_{2}$}, {HD}, and
  {D$_{2}$} from high-resolution spectroscopic data by multichannel
  quantum-defect theory},\ }\href {https://doi.org/10.1063/1.4866809}
  {\bibfield  {journal} {\bibinfo  {journal} {J. Chem. Phys.}\ }\textbf
  {\bibinfo {volume} {140}},\ \bibinfo {pages} {104303} (\bibinfo {year}
  {2014})}\BibitemShut {NoStop}%

\end{thebibliography}

\end{document}